\documentclass[10pt,preprint2]{aastex}
\usepackage{amsmath}
\usepackage{amssymb}
\usepackage{epsfig}

\newcommand{\cms}{\,{\rm cm$^{-2}$}\,}
\newcommand{\cmc}{\,{\rm cm$^{-3}$}\,}
\newcommand{\kms}{\,{\rm km\,s$^{-1}$}\,}

\newcommand{\kel}{\,{\rm K\ }}
\newcommand{\etal}{{ et~al.~}}

\newcommand{\ergs}{\,{\rm ergs\,s$^{-1}$}\,}
\newcommand{\ergscm}{\,{\rm ergs\,s$^{-1}$\,cm$^{-2}$}\,}
\newcommand{\Ms}{M_\odot}
\newcommand{\Zs}{Z_\odot}

\shorttitle{Nuclear Outflows in NGC~4552}
\begin{document}


\title{{\it Chandra} Observations of Nuclear Outflows in the 
        Elliptical Galaxy NGC~4552 in the Virgo Cluster}

\author{M. Machacek,  
 P.E.J. Nulsen$^\dagger$\altaffiltext{$\dagger$}{on leave from the
    University of Wollongong}, C. Jones, and W. R. Forman} 
\affil{Harvard-Smithsonian Center for Astrophysics \\ 
       60 Garden Street, Cambridge, MA 02138 USA
\email{mmachacek@cfa.harvard.edu}}

\begin{abstract}

We use a $54.4$\,ks {\it Chandra} observation  to 
study nuclear outflow activity 
in NGC~4552 (M89), an elliptical galaxy in the Virgo Cluster.
{\it Chandra} images in the 
$0.5-2$\,keV band show two ring-like features $\sim 1.7$\,kpc in 
diameter in the core of NGC~4552, as reported previously  
by Filho \etal  We use spherically symmetric  point explosion shock  
models to argue that the shape of the surface brightness profile 
across the rims of the rings and the temperature of hot gas in the 
rings are consistent with a Mach $1.7$ shock carrying 
mean mechanical power 
$L_{\rm shock} \sim 3 \times 10^{41}$\ergs 
produced by a $\sim 1.4 \times 10^{55}$\,ergs nuclear outburst 
$\sim 1 - 2 $\,Myr ago. We find the gas temperature 
in the central $\sim 100$\,pc of the galaxy to be $1.0 \pm 0.2$\,keV, 
hotter than elsewhere in the galaxy, suggesting that we may be
observing directly the reheating of the galaxy ISM by the outburst.

\end{abstract}



\section{INTRODUCTION}
\label{sec:introduction}

X-ray observations of nearby galaxies provide a unique opportunity 
to study directly the interaction of the central AGN with the galaxy's
surrounding hot gaseous atmosphere. {\it Chandra's} high angular resolution
permits the identification of outflow shocks, cavities, and 
buoyant bubbles that directly trace the flow of matter, energy and
momentum away from the nucleus and into the surrounding environment 
(e.g. see  Young \etal 2002, Forman \etal 2005 for M87 in the Virgo 
Cluster; Fabian \etal 2003 for NGC~1275 in the Perseus Cluster; 
McNamara \etal 2000, Nulsen \etal 2005 for Hydra A).

It is now widely recognized that understanding the nature of 
these AGN outflows is fundamental to further advances in the
understanding of the evolution of structure (galaxies, 
groups and clusters) and their hot gas environments, the interstellar 
(ISM) and intracluster (ICM) medium. 
Energy transfer to the ICM through episodic nuclear outbursts 
from a central supermassive black hole (Binney \& Tabor 1995; Owen 
\etal 2000) is likely necessary to resolve  
the order of magnitude discrepancy between the small amount of cool gas 
measured in the cores of rich clusters  
(e.g Tamura \etal 2001; Peterson \etal 2003 and references therein) and 
that predicted by standard cooling-flow models (Fabian 1994). 
Simulations demonstrated that the physical mechanisms responsible for 
the transfer of energy from the resultant relativistic plasma jets,  
shocks, buoyant bubbles, and sound waves to the surrounding 
gas are complex and strongly dependent on the age of the outburst
(e.g. Heinz \etal 1998, Churazov \etal 2001; Quilis \etal 2001; 
Br\"{u}ggen \& Kaiser 2002; Reynolds \etal 2001, 2005; Ruszkowski \etal
2004; Fabian \etal 2005; Heinz \& Churazov 2005; Omma \etal 2004).
 Young flows are characterized by shocks driven into the ambient medium 
by the outburst that heat the gas locally
(e.g. Heinz \etal 1998; Omma \etal 2004 and references therein). 
 After the outflow matures, fluctuations in the gas  
(``ripples'') passing over buoyant 
bubbles may induce Richtmyer-Meshkov instabilities that might 
dominate the transfer of bubble enthalpy to the ambient gas 
(Heinz \& Churazov 2005).
    
Recent studies indicate that feedback from a central AGN may be 
fundamental to our understanding of galaxy evolution in general.
Simulations, using the $\Lambda$CDM cosmology, suggest that nongravitational 
heating of the ISM by outbursts from the central, supermassive 
black hole in galaxies is needed to correctly reproduce 
the observed shape of the galaxy luminosity function and massive 
galaxy colors at low redshift (Granato \etal 2004; 
Croton \etal 2006 and references therein). 
Best \etal (2006) used FIRST and NVSS data to measure 
the fraction of galaxies that are 
radio-active above a given $1.4$\,GHz radio luminosity as a function
of that luminosity and central black hole mass. They interpreted their 
result as a probablistic estimate of the time a 
galaxy, with given black hole mass, is active above a given radio
brightness, and combined this result with the Birzan \etal (2004) 
correlation between the current central radio luminosity and 
the mechanical luminosity of outbursts 
to estimate the total energy available from AGN to heat the ambient
ISM. Best \etal (2006) argued that these data showed that 
heating from AGN could balance cooling in typical elliptical galaxies, 
and that most of the AGN heating in these  
galaxies occurs when the radio luminosity of the central AGN is low,
$L_{1.4\,{\rm GHz}} < 10^{31}$\ergs~Hz$^{-1}$.
Thus outflows may be important in  systems with low, as 
well as high, luminosity AGN.

In this work we use a $54.4$\,ks {\it Chandra} observation to study 
nuclear outburst activity in one such galaxy, NGC~4552 
 ($\alpha = 12^h35^m39.8^s$,$\delta =12^\circ33'23''$, J2000), an 
elliptical galaxy located $72'$ east of M87 
in subcluster A of the Virgo cluster.
In a companion paper 
(Machacek \etal 2006, hereafter called Paper I) we use the same 
{\it Chandra} observation to measure the properties
of hot gas in the outer regions of NGC~4552, showing that these
properties are consistent with that expected for gas being ram-pressure
stripped due to NGC~4552's supersonic 
motion through the Virgo ICM.
NGC~4552 is a member of the large 
sample of nearby elliptical galaxies used to establish correlations 
between X-ray emission and other ISM tracers, so that its global X-ray 
properties are well studied (for a brief review, see Paper I and 
references therein). 

Studies of the nuclear properties of NGC~4552, 
across a wide range of wavelengths, strongly 
suggest the presence of a weak AGN at its center. 
From optical line ratios, Ho \etal (1997) classify NGC~4552 as a 
LINER/HII transition object, possessing spectral properties of a 
low luminosity AGN contaminated by nearby star-forming HII regions. 
However, the classification was highly uncertain due to the weakness 
of the emission lines. HST spectra of the nuclear region 
by Cappellari \etal (1999) showed a faint broad emission component 
suggesting the presence of an AGN and favoring a LINER classification 
for NGC~4552's nucleus. Since the broad emission line component was 
observed in both the allowed and forbidden transitions, the nucleus 
was considered borderline between LINER types $1$ and $2$.
Short-term variability of the nucleus is a distinguishing feature of
AGN activity. UV variability has been reported for NGC~4552 
on timescales of several months (Maoz \etal 2005) to years 
(Renzini \etal 1995; Cappellari \etal 1999 ). Although not precluding 
a compact circumnuclear starburst, since variability in the UV 
on timescales $\lesssim 1$ year is not expected from young stars,   
 the variable flux provides a lower limit on the AGN contribution 
in this waveband (Maoz \etal 2005). 

Radio continuum measurements taken with the VLA at $8.4$\,GHz 
(Filho, Barthel \& Ho 2000) and with the VLBA at $5$\,GHz 
(Nagar \etal 2002) show a $\sim 100$\,mJy core radio 
source, unresolved on milli-arcsecond scales, at the galaxy
center. Variability at the $\sim 20\%$ level is observed in the $15$\,GHz and
$8.4$\,GHz radio flux on $\sim$~year timescales (Nagar \etal 2002).
Properties of the radio spectrum 
indicate a nonthermal origin for this radio emission, most likely 
associated with accretion onto a central supermassive black hole. The 
measured brightness temperature, 
$T_{\rm B} \sim 2 \times  10^{9}\kel$ at $5$ GHz (Nagar \etal 2002) 
is too high for this emission to be due to a nuclear starburst 
($T_{\rm B} \lesssim 10^4 -10^5\kel$) or collection of supernova 
remnants ($T_{\rm B} \lesssim 10^7\kel$). Furthermore, the shape of
the radio emission spectrum is too flat ($\alpha \sim 0$) to be 
associated with a collection of supernova remnants, whose 
spectral indices are typically  $\sim 0.4 -0.7$. Filho \etal (2004) 
attribute the flatness of the observed 
radio spectrum to ADAF-like accretion onto a central 
black hole combined with synchrotron emission from parsec scale jets 
or outflows. This is supported by the $5$\,GHz observation of two 
$\sim 5$\,milliarcsec diameter, extended features to the 
east and west of the core, suggestive of parsec scale jets 
(Nagar \etal 2002).  
From the correlation between the black hole mass and central velocity
dispersion (Ferrarese \& Merritt 2000; Gebhardt \etal 2000), 
the mass of the central black hole in NGC~4552 
is inferred to be $4 \times 10^8\Ms$.  In recent 
{\it Chandra} X-ray observations, Filho \etal (2004) identified  
a hard X-ray source, whose X-ray 
spectrum is well represented by a $\Gamma=1.51$ power law component 
plus a $kT=0.95$\,keV Raymond-Smith thermal component with
$0.5-2$\,keV and $2-10$\,keV luminosities of 
${\rm log}\,L_{\rm X} = 39.2$ and $39.4$ \ergs, respectively,  
that is coincident with the core radio source in NGC~4552. This    
is also consistent with the presence of a weak AGN.

Our discussion is organized as follows:  
In  \S\ref{sec:obs} we briefly review the observations and our 
data reduction and processing procedures.
In \S\ref{sec:ringresult} we  present the background 
subtracted, exposure corrected image of the central region of 
NGC~4552 showing two bright ringlike features within $~1.3$\,kpc of
the galaxy center. 
We model the spectral properties of these  
rings, in the central $1.3$\,kpc of the galaxy,   
and use the X-ray surface brightness profile to estimate 
the density jump across the rim of the rings. 
We show that these rings are
consistent with shocked gas driven outward by recent nuclear activity.
In section \S\ref{sec:nucleus}, 
the spectral properties of the nuclear region are briefly discussed 
for comparison with earlier work.
In \S\ref{sec:conclude} we summarize our results. 
Unless otherwise indicated, all errors correspond to  $90\%$ 
confidence levels and coordinates are J2000. Taking the distance to 
the dominant elliptical M87 as representative of the distance to 
subcluster A of the Virgo
Cluster containing NGC~4552, the luminosity distance to the cluster
is $16.1 \pm 1.1$\,Mpc (Tonry \etal 2001)  
and $1''$ corresponds to a distance scale of $77$\,pc. 


\section{OBSERVATIONS AND DATA REDUCTION}
\label{sec:obs}

We used a $54.4$\,ks observation of the Virgo elliptical galaxy
NGC~4552 taken with {\it Chandra} 
 on 2001 April 22 using the Advanced CCD Imaging Spectrometer array 
(ACIS, Garmire \etal 1992; Bautz \etal 1998) with ACIS-S (chip S3) at
the aimpoint. A detailed discussion of the observation and data 
analysis are presented in Paper I. We briefly review our procedures here.
The data were analyzed using the standard X-ray processing packages, 
CIAO version $3.1$, FTOOLS and XSPEC version $11.2$. 
Filtering removed events with 
bad grades ($1$, $5$, and $7$) and those with significant flux in 
border pixels of the  $5 \times 5$ event islands (VFAINT mode), as well
as any events falling on hot pixels.  
The data were reprocessed and response files created   
using the most recent gain tables and instrumental corrections, including 
corrections for the time-dependent declining efficiency of the 
ACIS detector due to the buildup of contaminants on the optical filter
(Plucinsky \etal 2003) 
and for the slow secular drift (tgain) of the average PHA values for photons
of fixed energy.\footnote{see Vikhlinin \etal in
http://cxc.harvard.edu/contrib/alexey/tgain/tgain.html}
Periods of anomalously high background (flares) were identified and
 removed, along with times of anomalously low 
count rates at the beginning and end of the
observation, resulting in a useful exposure time of $51,856$\,s. 
Point sources were identified in the 
S3 chip in the $0.3-10$\,keV energy band using a multiscale 
wavelet decomposition algorithm set with a $5\sigma$ detection
threshold. The resulting $132$ source identifications
were excluded from the spectral and surface brightness
analyses of NGC~4552 here and in Paper I.

\section{NUCLEAR OUTFLOWS IN NGC~4552}
\label{sec:ringresult}

\begin{figure*}[t]
\begin{center}
\includegraphics[height=2.4in,width=3in]{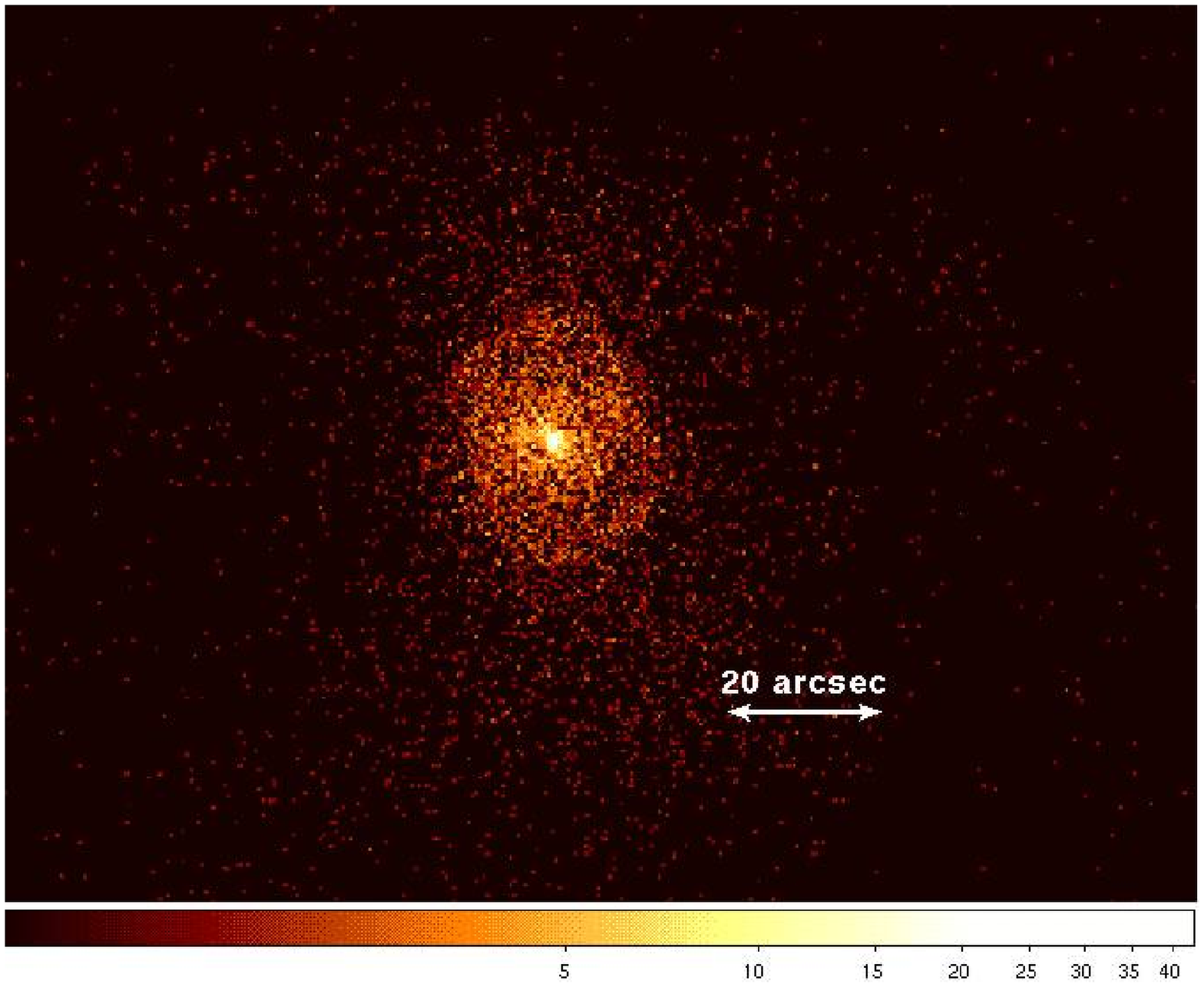}
\hspace{0.3cm}
\includegraphics[height=2.3in,width=3in]{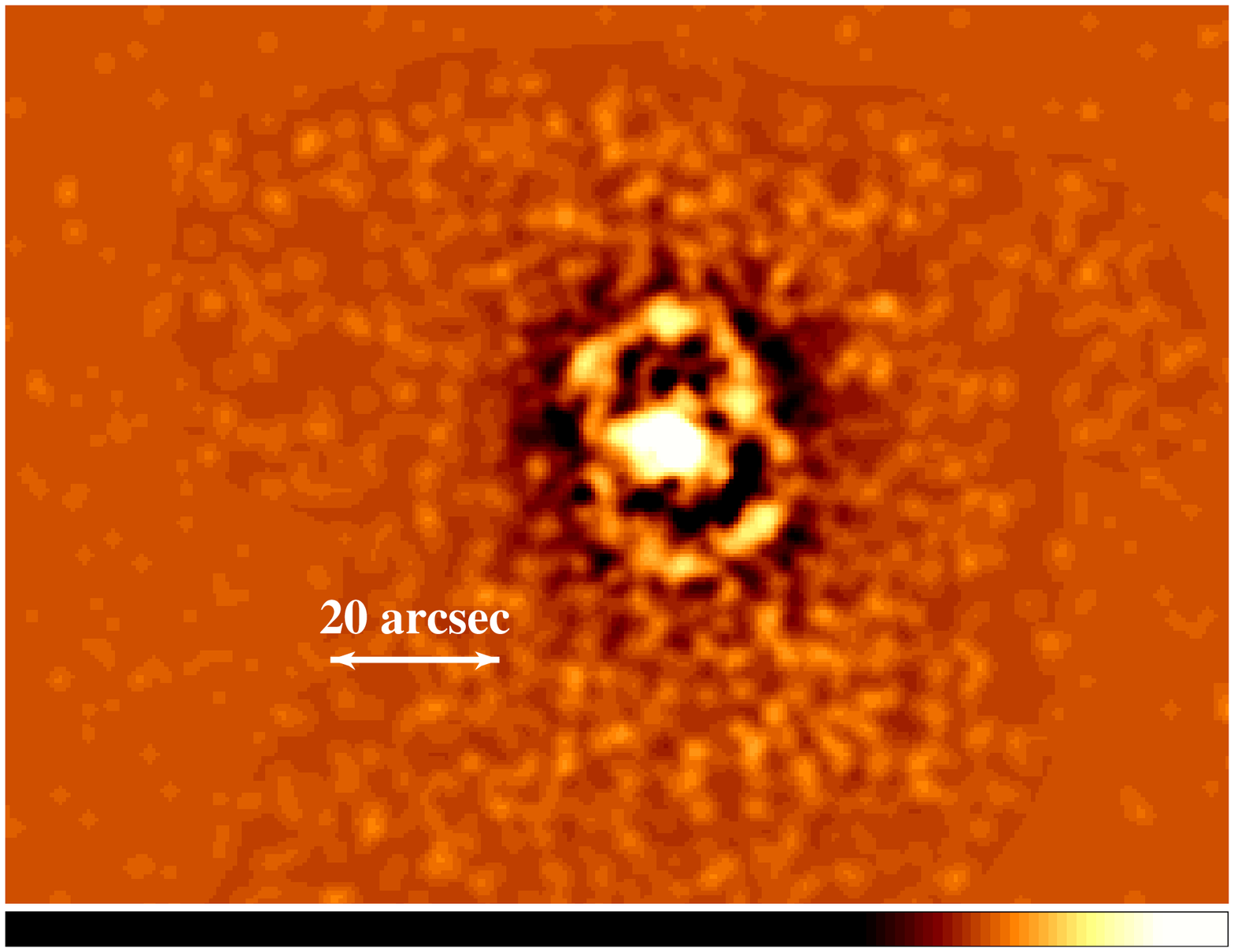}
\caption{\footnotesize{ (({\it left})  
{\it Chandra} image of the $0.7-1$\,keV diffuse emission from 
the inner rings in the central region of 
NGC~4552. $1\,{\rm pixel} = 0.''49 \times 0''.49$ 
({\it right}) Unsharp masked image of the same region, where 
we subtract the $0.7-1$\,keV diffuse image smoothed with a $5''.4$ Gaussian 
kernel from the same image smoothed with a 
$1''$ Gaussian kernel to highlight the rings.  
}}
\label{fig:ringimage}
\end{center}
\end{figure*}

In Figure \ref{fig:ringimage} we show the  
$0.7-1$\,keV diffuse emission from the two ring-like features in the 
central region of NGC~4552. Point sources were removed from the 
image and the point source regions filled with the local 
average emission level using CIAO tool {\it dmfilth}. 
In the right panel of Figure \ref{fig:ringimage} we subtract the
$0.7-1$\,keV image smoothed with a $5''.4$ Gaussian kernel from the 
same image smoothed with a $1''$ Gaussian kernel to produce an
unsharp masked image of the same region, in which these rings 
are clearly visible.
The rings are approximately circular and of equal size 
with radii $\sim 0.85$\,kpc ($11''$) and farthest projected distance 
from the nucleus of NGC~4552 of $\sim 1.3$\,kpc ($\sim 17''$), creating
in projection the 'hourglass' appearance noted by 
Filho \etal (2004). Such cavities have
been shown in other elliptical galaxies, such as M87 (Young \etal
2002; Forman \etal 2005), to chronicle past episodes of nuclear 
activity from a central AGN. 

\subsection{Gas Temperature in the Rings}
\label{sec:ringtemp}

\begin{figure}[t]
\begin{center}
\includegraphics[height=2.65in,width=3in]{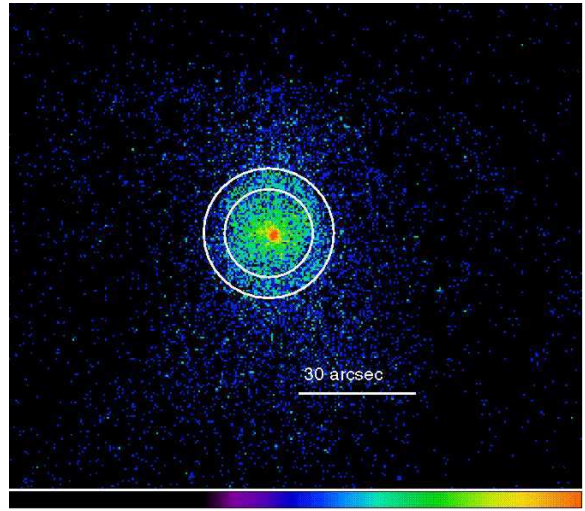}
\caption{\footnotesize{ Background subtracted 
{\it Chandra} image of the $0.5-2$\,keV diffuse emission from 
the inner rings in the central region of 
NGC~4552 with spectral region R
superposed. $1\,{\rm pixel} = 0.''49 \times 0''.49$.
}}
\label{fig:ringspecreg}
\end{center}
\end{figure}

We fit the spectrum of the rings in an annular region R, shown in 
Figure \ref{fig:ringspecreg} and listed 
in Table \ref{tab:centralregs}, with
inner and outer radii of $11''$ and $17''$, respectively. The local 
background region for the rings (also listed in Table \ref{tab:centralregs})
was chosen concentric to and immediately outside the source to
remove contamination from galaxy emission at larger radii as well as the
Virgo ICM. We use an absorbed APEC thermal plasma model (Smith \etal
2001) corrected for absorption using Wisconsin photo-electric
cross-sections (Morrison \& McCammon 1983), with the 
absorbing column fixed at the Galactic value of $2.59 \times
10^{20}$\cms (Dickey \& Lockman 1990), to fit the 
 resulting $1840$ net source counts over the $0.3-3$\,keV energy range,
where the source count rate was significantly above background. We 
 allowed the hydrogen column to vary and found no suggestion for 
increased absorption above Galactic. 
We checked that there were no significant differences between 
the spectra for the northern and southern rings by fitting the northern
and southern halves of region R separately. We found the temperature 
and abundance of the northern (southern) ring to be $0.64 \pm 0.03$\,keV 
($0.65^{+0.03}_{-0.04}$\,keV) and $0.6^{+0.4}_{-0.3}\,\Zs$ 
($ > 0.5\,\Zs)$, respectively, for fixed Galactic absorption. 
Since the spectral fits for the northern and southern rings agreed 
within errors, we combined these regions into the single annular 
region R to improve statistics.

The best fit APEC model spectral parameters for gas in the rings 
from this combined annular region (Region R) are listed in the 
first row of Table \ref{tab:ringfits}. 
We find the temperature and abundance of gas in
the rings to be $kT = 0.64 \pm 0.02$\,keV and 
$A = 0.7^{+0.4}_{-0.2}\,\Zs$ ($\chi^2/{\rm dof} = 46/55$), respectively.
Thus the temperature of gas in the rings (region R) is a factor $1.4$ 
 higher than that found for the surrounding
 ($0.43^{+0.03}_{-0.02}$\,keV) galaxy gas (Paper I). This is  
in contrast to the properties of a cold front, where the brighter
region is cooler than its less bright surroundings (Viklinin \etal
2001; see also Paper I and references therein) or to the cool, bright 
rims of  highly evolved remnant cavities 
from nuclear activity, such as those found in Perseus (Fabian \etal
2003) and in M87 (Forman \etal
2005), but is similar to the properties observed in the X-ray edges
produced by shocks driven into the ambient medium from recent 
AGN outbursts, such as in Hydra A (Nulsen \etal 2005a, 2005b).

We verified that the  temperature rise in the rings   
cannot be due to unresolved X-ray 
binaries in our data by adding a $5$\,keV bremsstrahlung component 
to our spectral model, as in Paper I, and refitting the
spectrum over the $0.3-5$\,keV energy range.  The resulting best fit 
temperature ($0.63^{+0.03}_{-0.02}$\,keV) for the thermal component in this 
two component model agrees with the single component 
APEC model results to within $2\%$. The abundance of the gas 
($A=0.9^{+0.5}_{-0.3}\,\Zs$) is also in agreement, 
within the $90\%$\,CL model uncertainties, with that found using a  
single APEC model. The $0.5-2$\,keV X-ray luminosity of region R is 
dominated by emission from galaxy gas, with the bremsstrahlung 
(unresolved X-ray binary) component contributing $< 5\%$ to the 
$0.5-2$\,keV luminosity in that region. This is in agreement 
with expectations for the residual contribution of 
unresolved X-ray binaries obtained 
from a direct integration of the average XLF for LMXB's 
(Gilfanov, 2004) below our $0.5-8$\,keV point source detection 
threshold ($4.8 \times 10^{-15}$\ergscm) for region R. Thus 
the effects of unresolved X-ray binaries on the measurement of 
the properties of the diffuse gas in the rings that are 
important for our analysis are small.

Although the uncertainties are large, the metallicity in the rings is
consistent with that found in Paper I for the outer regions of the galaxy. 
Thus we find no evidence for a strong metallicity gradient across 
the rings. For completeness we list in Table \ref{tab:ringfits} 
the metallicity dependence of the $0.5-2$\,keV
X-ray emissivity for gas in the rings over the abundance range 
($0.4 -1.0\,\Zs$) of interest for NGC~4552.

\subsection{Gas Density in the Rings}
\label{sec:ringdens}

\begin{figure}[t]
\begin{center}
\includegraphics[height=3in,width=3in,angle=270]{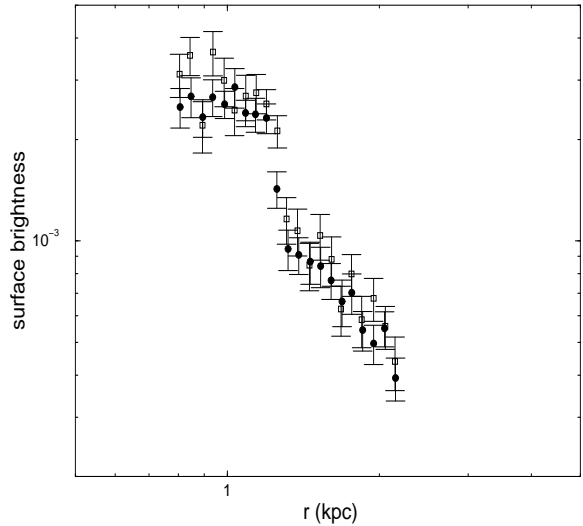}
\caption{\footnotesize{The $0.5-2$\,keV surface brightness profiles across the
northern (open squares) and southern (filled circles) rings taken 
in angular sectors, whose vertices are positioned at the  
nucleus and angles extend from $64^\circ$ to $121^\circ$ for the northern 
rim and from $257^\circ$ to $349^\circ$ for the southern rim 
(measured counterclockwise from the west in 
Figure \protect\ref{fig:ringimage}) 
 }}
\label{fig:ringsbcomp}
\end{center}
\end{figure}

\begin{figure*}[t]
\begin{center}
\includegraphics[height=3in,width=3in,angle=270]{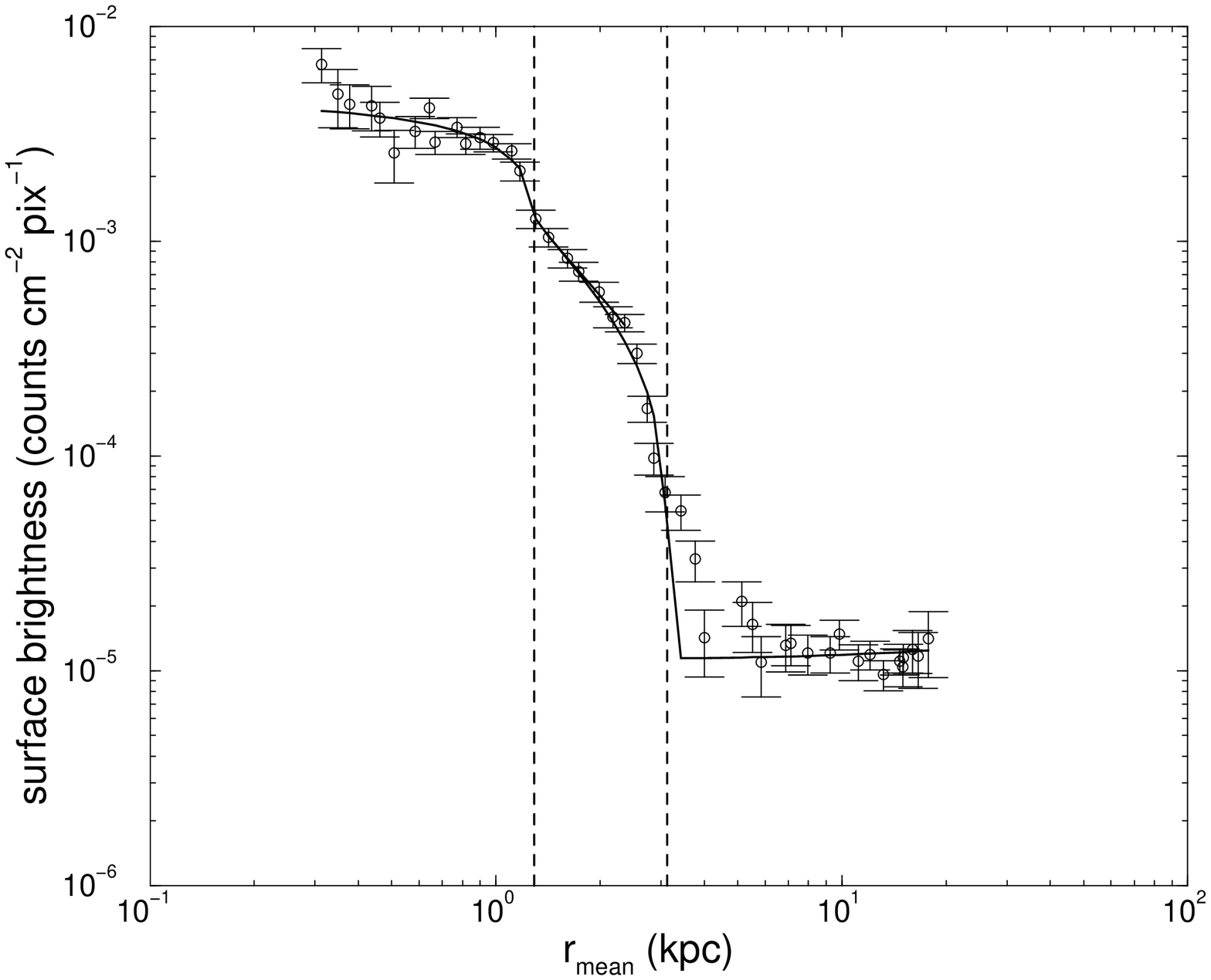}
\hspace{0.1cm}
\includegraphics[height=3in,width=3in,angle=270]{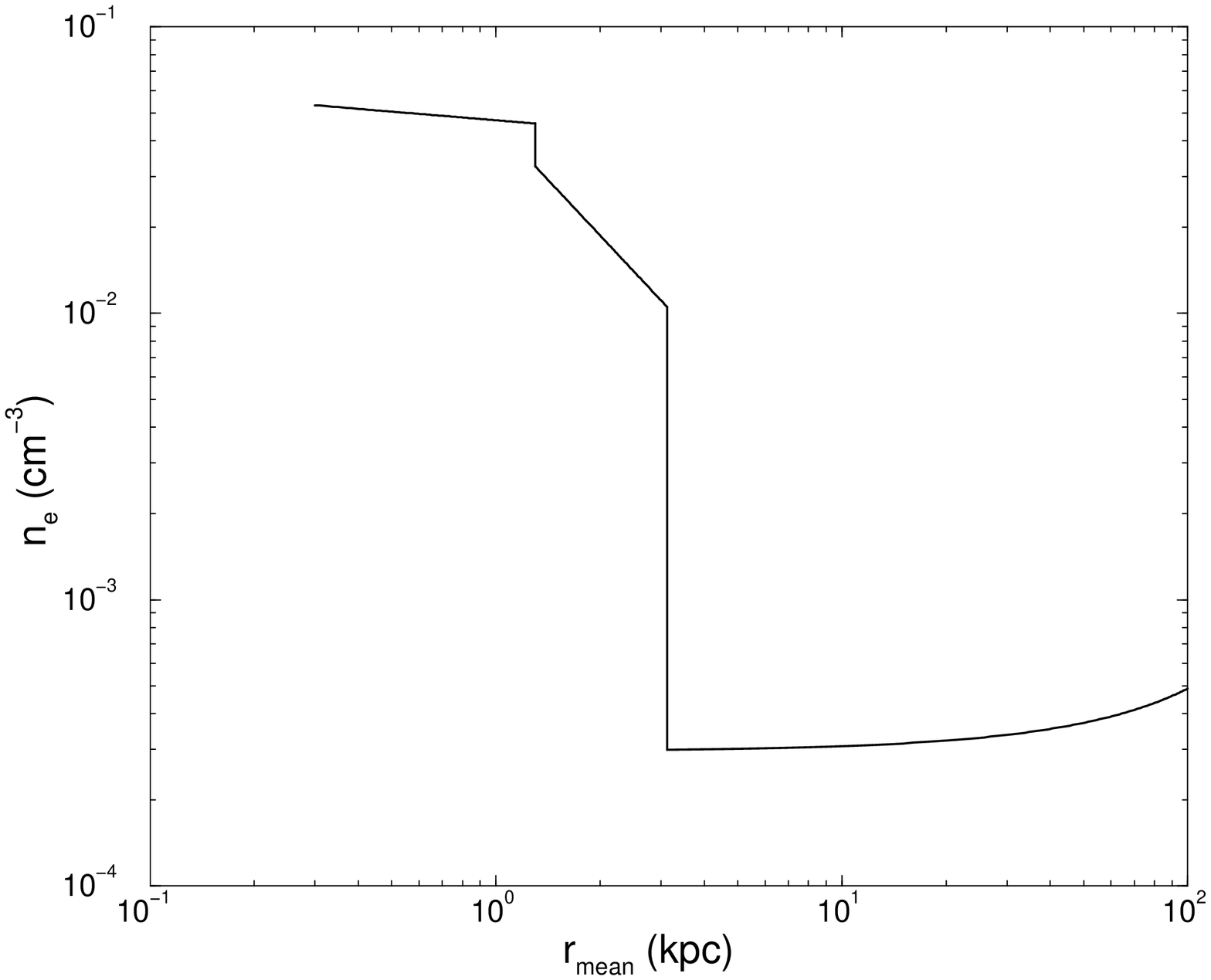}
\caption{\footnotesize{(left) The $0.5-2$\,keV surface brightness profile 
as a function of distance from the center of NGC~4552 toward the north
across the galaxy's leading edge from Paper I.  
The vertical dashed lines denote $r_1=1.3$\,kpc and $r_2=3.1$\,kpc, 
the position of the outer radius (rim) 
of the northern ring and the outer leading edge of the cold 
front, respectively. 
The solid line denotes the 
model results for the surface brightness for power law density 
models $n_e \propto r^{-\alpha}$ within the galaxy with
$\alpha = 0.1^{+0.25}_{-0.1}$ for $r <1.3$\,kpc 
(inside the rim of the ring) and $\alpha = 1.3^{+0.2}_{-0.2}$
for $1.3 \le r < 3.1$\,kpc (between the rim of the ring 
and the leading edge of the cold front).  
(right) The electron density as a
function of radius from the center of NGC~4552 for the above model.
}}  
\label{fig:sbprof}
\end{center}
\end{figure*}

As shown in Figure \ref{fig:ringsbcomp}, the $0.5- 2$\,keV
surface brightness profiles across the northern and 
southern rims, taken in an angular sector centered on the 
nucleus extending from $64^\circ$ to $121^\circ$ for the northern 
rim and $257^\circ$ to $349^\circ$ for the southern rim 
(measured counterclockwise from west in Figure \ref{fig:ringimage}),
agree within errors. Thus we use the surface brightness profile
across the northern rim as representative of the ring system.
We use the multivariate minimization technique (Markevitch \etal 2000;
Vikhlinin \etal 2001) to determine the electron density inside 
the rim of the northern ring from the surface brightness profile 
shown in Figure \ref{fig:sbprof}.

We assume a spherically symmetric power law distribution for 
the electron density for $r \leq r_1$ 
of the form  
\begin{equation}
n_{\rm e} = n_{\rm rim}\Bigl (\frac{r}{r_1} \Bigr )^{-\alpha_1}, 
\label{eq:galinner}
\end{equation}
where $r_1$ is the outer radius(rim) of the ring, and $n_{\rm rim}$
and $\alpha_1$ are the normalization and power law index,
respectively, for the electron density of NGC~4552 inside the rim 
of the northern ring. 
The density model for NGC~4552's gas outside the ring $r_1 \le r \le
r_2$, determined by our analysis of the northern leading edge in Paper
I, is given by 
\begin{equation}
n_{\rm e} = n_{{\rm e}2}\Bigl (\frac{r}{r_2} \Bigr )^{-\alpha_2}\,,
\label{eq:galouter}
\end{equation}
 where $r_2 = 3.1$\,kpc, $\alpha_2=1.3 \pm 0.2$, and 
$n_{{\rm e}2}=0.01$\cmc.
Since NGC~4552  is much  
smaller than the projected $71'.6$ separation between M87 and NGC~4552, the
contribution of the cluster ICM to the surface brightness for radii 
$r \le 3.1$\,kpc (inside the galaxy) is nearly constant, varying 
by $\lesssim 2\%$.
We subtract this 
ICM contribution from the surface brightness profile within NGC~4552 
($r < 3.1$\,kpc) and fit the remaining galaxy emission, 
taking the position $r_{\rm rim}$ of the rim of the ring,
the galaxy electron density power law indices $\alpha_1$, $\alpha_2$, 
and the discontinuity across the rim as free parameters. The
discontinuity across the rim is given by 
\begin{equation}
 J = \Bigl (\frac{\Lambda_{\rm rim} n_{\rm rim}^2}{\Lambda_{\rm out}n_{\rm out}^2} \Bigr ),
\label{eq:rimjump} 
\end{equation} 
where   $n_{\rm  rim}$ and $\Lambda_{\rm rim}$  
denote the electron density and 
emissivity inside the rim of the northern ring, and
$n_{\rm out}$ and $\Lambda_{\rm out}$ denote the corresponding properties of 
galaxy gas outside the rim.

We find a best fit position for the discontinuity across the rim shown in 
Figure \ref{fig:sbprof} of $r_1=1.30$\,kpc, 
coincident with the rim (outer radius) of the northern ring in Figure
\ref{fig:ringimage}, 
and slope $\alpha_1 = 0.1 \pm 0.3$ for the electron density at $r \leq
r_1$. The slope of the electron density distribution for radii 
$r > r_1$ (outside the rings) was  $\alpha_2 =1.3 \pm 0.2$, 
consistent with our previous fit in Paper I for the galaxy gas at large radii.
Assuming no strong abundance gradients in NGC~4552 and correcting for the 
$20\%$ difference in the galaxy gas cooling functions, caused
by the difference in gas temperature between gas at large radii
($0.43$\,keV, Paper I) and gas in the rings ($0.64$\,keV
for region R), we infer the density ratio across the rim of the
northern ring to be  $n_{\rm rim}/n_{\rm out} = 1.4^{+0.3}_{-0.2}$.
We extrapolate eq. \ref{eq:galouter} to $r=1.3$\,kpc, the location
of the rim of the northern ring, to evaluate $n_{\rm out}$ at the
discontinuity. Then using the density discontinuity
at the rim, we find the electron density just inside the rim 
to be $n_{\rm rim} = 0.05$\cmc.
In the right panel of Figure \ref{fig:sbprof}, we plot  
the resulting normalized two power law density model for gas in 
NGC~4552 and the $\beta$-model for the surrounding
Virgo ICM as a function of distance from the center of NGC~4552.

\subsection{Shocks from Recent Nuclear Activity}
\label{sec:shock}

Since the temperature ($kT_{\rm rim} =0.64$\,keV) and density 
($n_{\rm rim} = 0.05$\cmc) 
of gas in the inner rings are both greater than that for galaxy gas 
outside the rings ($kT_{\rm out} \sim 0.43$\,keV, 
$n_{\rm out} \sim 0.03$\cmc) at 
$r_1=1.3$\,kpc, the properties of the rings are 
qualitatively consistent with resulting from a shock. 
For a monatomic ideal gas with adiabatic index $\gamma$, we can use
the density discontinuity 
$\rho_{\rm rim}/\rho_{\rm out} = n_{\rm rim}/n_{\rm out}$ 
in the Rankine-Hugoniot shock conditions 
(Landau \& Lifschitz 1959) across the shock to estimate the shock 
speed (Mach number $M_1$)
\begin{equation}
 \frac{n_{\rm rim}}{n_{\rm out}} = 
  \frac{(\gamma + 1)M_1^2}{(\gamma -1)M_1^2 + 2}
\label{eq:shockd} 
\end{equation}
and predict discontinuities in temperature
\begin{equation}
 \frac{T_{\rm rim}}{T_{\rm out}} = 
  \frac{(2\gamma M_1^2-(\gamma - 1))((\gamma -1)M_1^2 + 2)}{(\gamma + 1)^2M_1^2}
\label{eq:shockt} 
\end{equation}
and pressure
\begin{equation}
 \frac{p_{\rm rim}}{p_{\rm out}} = 
  \frac{2\gamma M_1^2}{\gamma + 1}-\frac{\gamma -1}{\gamma + 1}\,.
\label{eq:shockp} 
\end{equation}
We can obtain an estimate for the density jump 
$n_{\rm rim}/n_{\rm out}$ from the 
surface brightness discontinuity at the position of the rings.
However, since the shock front is narrow, the averaged, 
measured surface brightness discontinuity will underestimate the 
actual density discontinuity at the narrow shock front, and 
thus also the inferred shock strength and 
temperature rise.
From our fit to the inner surface brightness edge in 
\S\ref{sec:ringdens}, we infer a lower limit on 
the density discontinuity of
$1.4^{+0.3}_{-0.2}$. From Equation \ref{eq:shockd} this implies a 
shock Mach number $M_1 \gtrsim  1.3$. For this Mach number, 
the temperature and
pressure increases  predicted by Equations \ref{eq:shockt} and
\ref{eq:shockp} are $T_{\rm rim}/T_{\rm out} \gtrsim 1.3$ and
$p_{\rm rim}/p_{\rm out} \gtrsim  1.8$, 
in excellent agreement
with our measured ratios of  $1.48^{+0.15}_{-0.13}$ and 
$2.1^{+0.7}_{-0.5}$, respectively. 
Thus the bright rings in Figure \ref{fig:ringimage} are  
shocks at the interface between the surrounding galaxy gas and two  
expanding cavities. Numerical simulations have predicted that such 
features would be formed by high-momentum, subrelativistic bipolar nuclear 
outflows (see e.g. Reynolds \etal 2001; Omma \etal 2004 and references
therein). 
 
An estimate for the age of  
these cavities is given by $t_{\rm age} \sim R/c_s$ where $R$ is the 
mean radius of the cavity and $c_s$ is the sound speed in the ambient
medium (Birzan \etal 2004). For $R \sim 0.85$\,kpc and 
$c_s \sim 340$\kms, the speed of sound in $0.43$\,keV galaxy gas, 
we find the cavities to be $\sim 2.4$\,Myr old. 
A lower limit on the total mechanical energy $E_{\rm mech}$ carried in 
the outflow is given by the work $2p_gV$ needed to evacuate the 
two observed cavities, where $p_g$ is the mean pressure of the 
unshocked galaxy gas at each cavity's center
and $V$ is the volume of the cavity.  Assuming spherical symmetry 
for each expanding cavity and extrapolating the  density model 
in Equation \ref{eq:galouter} for unshocked galaxy
gas  outside the rings to the center of each cavity, we find 
the sum of the mechanical energy for the two cavities to be 
$E_{\rm mech} > 1.2 \times 10^{55}$\,ergs. 
This gives a lower limit on the average mechanical power for the 
outburst of $L_{\rm mech} > 1.5 \times 10^{41}$\ergs. 

\begin{figure*}[t]
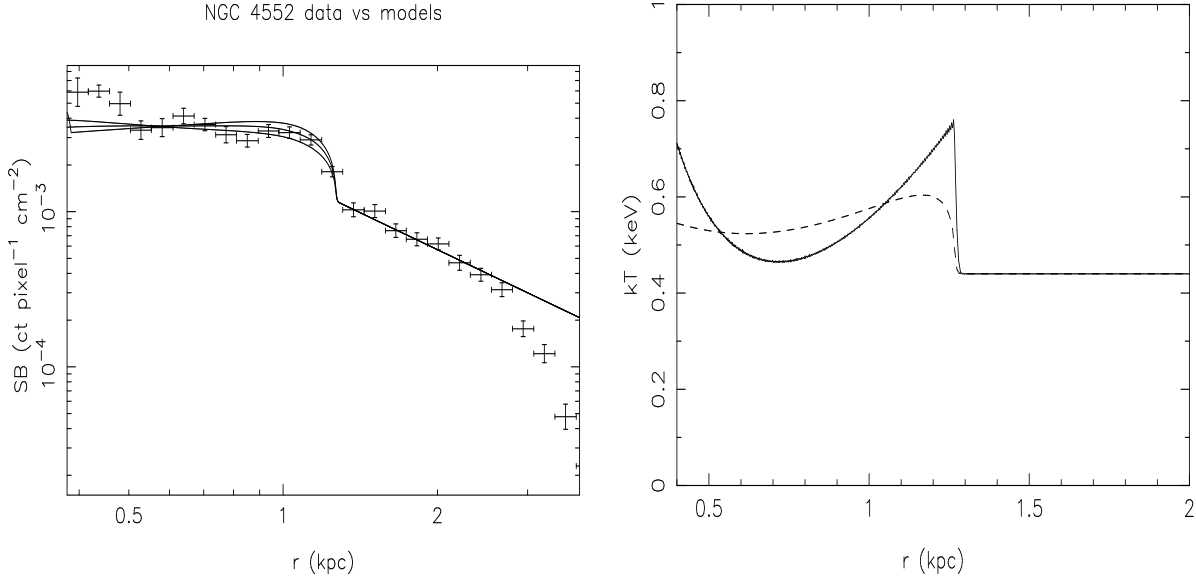

\begin{center}
\includegraphics[height=3in,width=3in,angle=270]{f8a.ps}
\hspace{0.3cm}
\includegraphics[height=3in,width=3in,angle=270]{f8b.ps}
\caption{\footnotesize{ ({\it left}) Spherical shock model fits to the surface
brightness profile across the inner edge for 
a preshocked isothermal ($0.43$\,keV) $r^{-1.3}$ density distribution
inside the galaxy. Upper, middle and lower
lines represent shock Mach numbers 1.9, 1.7 and 1.6, respectively. 
({\it right}) Temperature profile (solid line) for the Mach 1.7 model
shock shown in the left panel. Dashed line denotes the
emission-measure-weighted temperature profile for lines of sight 
through the galaxy.
}}
\label{fig:shckmod}
\end{center}
\end{figure*}

We use a simple, spherically symmetric, point explosion shock model 
to obtain a second estimate for the properties of the outflow. 
The details of this model and its uncertainties are discussed in 
Nulsen \etal (2005a, 2005b). We briefly summarize that discussion 
here. The model shock is produced by a sudden deposition of 
energy in the central cell of the simulation. The model then 
uses gas hydrodynamical equations, including gravity, to numerically 
propagate the shock radially outward through the galaxy gas, assuming 
the unshocked galaxy atmosphere is isothermal, spherically symmetric 
with a power law density distribution, and in hydrostatic equilibrium.
Assuming the gas in NGC~4552 is in hydrostatic equilibrium, we use our 
fits to the outer region (OA in Paper I) of NGC~4552 as 
representative of the temperature ($kT=0.43$\,keV) 
and density distribution ($\propto r^{-1.3}$) of the 
initially unshocked galaxy gas. We then compute surface brightness
profiles from the model, using the {\it Chandra} response in the 
$0.5-2$\,keV energy band, determined using XSPEC and 
an absorbed APEC model for temperatures determined from the shock
model, assuming a preshock temperature of $kT=0.43$\,keV 
and detector responses appropriate to this observation. 
Since the shock weakens as
it hydrodynamically evolves, and the initial conditions are
self-similar, we scale the model flow radially to the position of the 
observed shock. We then match the observed surface brightness profile to
that of the model to determine the shock parameters. 
Although this point explosion model will not reproduce the detailed
(bipolar) geometry of the outflows, the model does provide a good 
description for the shape of the surface brightness profile across 
the rim and thus a reasonable measure of the Mach number, and thus  
temperature jump, of the shock (Nulsen \etal 2005b). 

In the left panel of Figure \ref{fig:shckmod} we see that 
the surface brightness profile across the rim of the northern ring 
in  NGC~4552 is well represented by a shock model with 
Mach number $M_1 = 1.7^{+0.2}_{-0.1}$ and shock radius of $1.29$\,kpc.
The predicted, deprojected temperature profile across the shock front for the 
$M_1 = 1.7$ model
is shown as the solid line in the right panel of 
Figure \ref{fig:shckmod}. The radial 
extent of region R ($0.9 \lesssim r \lesssim 1.29$\,kpc), used to 
measure the temperature inside the rings, is much larger than the 
radial width of the shock. Our measured temperature (
$0.64 \pm 0.02$\,keV from Table \ref{tab:ringfits}) represents  
an average over this region and is in good agreement with the model 
results.  Projection effects also tend to reduce 
the observed temperature, since any given line of sight through 
the shock also passes through surrounding unshocked galaxy gas of 
lower temperature. The dashed line in the right panel shows 
the predicted emission-measure-weighted temperature profile $T_{\rm proj}$ 
for our Mach $1.7$ model, taking into account the contribution from 
cooler preshocked galaxy gas along the line of sight. The 
emission-measure-weighted temperature profile rises sharply at the 
shock boundary, peaking at $\sim 0.6$\,keV, and then slowly decreases 
to $\sim 0.55$\,keV at $r =0.9$\,kpc, the inner radius of region R.
This predicted profile is shifted up or down by $\sim 0.03$\,keV 
due to uncertainties in the temperature of the preshocked galaxy gas. 
Since the model $T_{\rm proj}$ includes the contribution of 
unshocked galaxy gas at radii larger than that of the rings, that we 
subtracted as a local background in our previous fits to region R, we 
refit the spectrum for the rings in region R, for comparison with the
model, using a $6'.7 \times 2'.5$ rectangular region (region VN in 
Paper I) centered at ($\alpha =12^h35^m37.7^s$,
$\delta =12^{\circ}36'16.5''$) with orientation angle of $319^{\circ}$, 
comprising approximately $25\%$ of the area of 
the S3 chip, as the local background, to subtract only the 
Virgo ICM and particle backgrounds from the data.
We find a mean temperature $kT_{\rm proj} = 0.61 \pm 0.02$\,keV and 
abundance $A=0.6 \pm 0.15$ ($\chi^2/{\rm dof} = 51.8/55$)
for fixed Galactic absorption. This is in excellent agreement with the shock 
model prediction (Figure \ref{fig:shckmod}, just behind the shock). 
We verified that this result was not biased by our 
choice of local background by refitting the spectrum 
using blank sky backgrounds appropriate for the date of observation
and instrument configuration.\footnote{see 
http://cxc.harvard.edu/contrib/maxim/acisbg and Paper I} 
When we used a second APEC component to model the contribution of the Virgo 
cluster gas, with temperature $2.2$\,keV and abundance $0.1\Zs$ fixed
at  the ``best-fit'' Virgo Cluster values from Paper I, 
but with the normalization free to vary,  we found no significant 
differences.

The bipolar geometry of the outburst cavities, seen most clearly in 
the right panel of Figure \ref{fig:ringimage}, makes it difficult to 
probe the temperature profile inside the rims. To investigate this 
region, we extract a combined spectrum from elliptical regions in the 
northern (southern) ring  
centered at $\alpha = 12^h35^m39.87^s$,$\delta=+12^\circ 33'31.7''$ 
($\alpha = 12^h35^m39.53^s$,$\delta = +12^\circ 33'15.97''$) with 
semi-major, minor axes and position angle (a, b, pa) of 
$a= 0.47$\,kpc,$b=0.27$\,kpc,${\rm pa}=9.5^\circ$ 
($a=0.52$\,kpc,$b=0.20$\,kpc,${\rm pa}=21.3^\circ$), respectively. 
We use these elliptical regions rather than a simple annular region 
inside the rings in order to avoid the region just outside the nucleus 
where, in projection, the bipolar ring rims appear to intertwine. We model 
the spectrum in XSPEC with an absorbed two component APEC model,
fixing the temperature and abundance of one APEC component at the 
the temperature and abundance of the Virgo Cluster gas, as above, 
and hydrogen absorption at the Galactic value. We find a mean temperature
and abundance for galaxy gas averaged along the lines of sight passing 
through the interior of the cavities of 
$kT_{\rm proj}=0.65^{+0.03}_{-0.04}$\,keV and 
$A=0.6^{+0.4}_{-0.2}\Zs$ ($\chi^2/{\rm dof} = 43/34$), respectively. 
This agrees within $90\%$ 
confidence with the mean temperature and abundance 
($kT_{\rm proj,rim}= 0.61 \pm 0.02$\,keV, $A=0.6 \pm 0.15\Zs$) 
measured in the annular ring region R. 

Thus we do not observe the 
$\lesssim 10\%$ decline in temperature behind the shock that might 
have been expected from the model temperature profile in the right
panel of Figure \ref{fig:shckmod}. This is not surprising for several 
reasons. First, both the ring and cavity spectral regions average 
over a broad range in radius. Region R, with inner and outer radii of 
$0.9$ and $1.3$\,kpc, averages over $30\%$ of the radial distance from 
the shock rim to the nucleus, thus already averaging over the region 
where the temperature decline due to adiabatic expansion of gas 
behind the shock front should be greatest.
Similarly the elliptical regions for the cavities also 
sample a broad range in radius, that is further complicated by
projection effects due to the bipolar geometry of the outflows.
Secondly, the model evolution of the gas temperature behind the shock front 
is sensitive to the density and temperature profile of the unshocked 
atmosphere. The unshocked gas density in galaxies typically follows a
$\beta$-model profile, flattening in the core. Thus it is strictly a 
power law only locally near the rim. Also  galaxy gas
near the nucleus of NGC~4552  may not be isothermal. Finally,
the outflows are likely still driven by the ram-pressure of jet plasma 
from the initial AGN outburst, that has effectively evacuated X-ray
gas from a 'cocoon' inside the shocked rims, rather than by a point 
explosion model. This would also affect the observed thermal properties of 
gas averaged over lines of sight passing through the 'cocoon'. 
In particular, if the cavities are nearly devoid of X-ray gas, the 
emission will be dominated by the front and back edges of the cavities. 
Additional discussion of these model uncertainties can be found
in Nulsen \etal (2005b).

Nonetheless, the model does fit the surface brightness profile in the 
vicinity of the shock well (see Fig. \ref{fig:shckmod}), such that the 
shock Mach number is largely insensitive to the details of the model. 
Consequently, the point explosion model estimates for the total 
mechanical energy and mean mechanical power carried in such outbursts 
are unlikely to err from their true values by more than a factor 
$\sim 2$ (Nulsen \etal 2005b).  Adopting a factor two as a conservative
estimate for these model uncertainties, we now use the shock model fit
to our data to determine the  total mechanical energy and 
mean mechanical power required for the nuclear outburst observed in NGC~4552. 
The elapsed time since the 
nuclear outburst (shock age) in the Mach $1.7$ spherical 
shock model  is $1.3$\,Myr. Since the simplified spherical 
model geometry overestimates the volume swept out by the observed 
shocks, the model predictions for the total mechanical energy 
$E_{\rm mech}$ carried in the shocks and thus also the mean mechanical 
power $L_{\rm shock}$ 
 of the outburst are upper bounds on the
properties of the observed outburst in NGC~4552. We correct for this 
by including a geometrical factor $f_{\rm V} \le 1$, 
where $f_{\rm V}$ is approximately the ratio of the volume filled by the
observed cavities to that of the spherical model. 
For a Mach $1.7$ shock, we find 
$E_{\rm mech} = 2.4 \times 10^{55}f_{\rm V}$\,ergs and
$L_{\rm shock} = 5.8 \times 10^{41}f_{\rm V}$\ergs.  
Assuming the observed cavities are well approximated by 
two spheres of radius $R=0.85$\,kpc, we find the fraction of the model 
volume filled by these cavitities is $f_{\rm V} \sim 0.6$,  such that  
the total mechanical energy and mean mechanical luminosity of the outburst
are $E_{\rm mech} \sim  1.4 \times 10^{55}$\,ergs and 
$L_{\rm shock} \sim 3.3 \times 10^{41}$\ergs, respectively, 
consistent with our previous estimate. 

We obtain an estimate of the total X-ray cooling rate for the galaxy 
ISM by fitting a mean spectrum for NGC~4552 extracted from a 
$50''$ ($3.8$\,kpc) circular region centered on the galaxy's nucleus,
using a two component absorbed APEC model with absorption fixed at 
Galactic and one set of parameters fixed at the best fit Virgo Cluster
parameters. We find a mean temperature ($0.57 \pm 0.01$\,keV), 
abundance ($0.5 \pm 0.1\,\Zs$), and $0.2-2$\,keV flux of 
$F(0.2-2) = 7 \times 10^{-13}$\ergscm for NGC~4552's ISM 
component, in excellent agreement with results 
($kT = 0.55^{+0.11}_{-0.08}$\,keV, 
$F(0.2-2) = 6.8^{+1.1}_{-1.0} \times 10^{-13}$\ergscm) by 
Matsushita (2001) using ROSAT data.  Using the $0.1 - 14$\,keV
luminosity as an estimate for the bolometric luminosity 
$L_{\rm bol}$ of the hot gas, 
we find $L_{\rm bol} = 2.5 \times 10^{40}$\ergs. If we use a more realistic 
two component temperature model for the ISM, consistent with our 
spectral results for the individual regions, this  estimate of the 
total flux decreases by $\lesssim 2\%$. Therefore the energy carried in the 
observed nuclear outflows ($E \sim 1.4 \times 10^{55}$\,ergs) is 
sufficient to balance cooling of the galaxy ISM for $\approx 18$\,Myr.

Energetic outbursts could either be powered by supernovae from 
a compact circumnuclear region of star formation, i.e. 
in the central $\sim 100$ pc ($1''.3$ radius) of the galaxy, or 
by an AGN at the galaxy's center. Assuming $10^{51}$\,ergs kinetic 
energy per supernova and $\sim 1$ supernova per $100\,\Ms$ of star 
formation (McNamara\etal 2004; Silk \etal 1986), we would need 
$\gtrsim 14,000$ supernova 
to provide the outburst energy of $1.4 \times 10^{55}$\,ergs 
(for $f_{\rm V} = 0.6$). This would 
require a mean star formation rate, over the $1.3$\,Myr age of the
outburst, of $\gtrsim 1\,\Ms\,{\rm yr}^{-1}$ within $100$\,pc of the 
nucleus. If we make the extreme assumption 
that all star formation in NGC~4552 occurs in this circumnuclear region, 
we can use the total UV emission of the galaxy at $2500$\AA\,       
as a direct tracer of light from unabsorbed star formation, and the 
TIR emission ( IR emission integrated over the $8 - 1000\,\micron$ band) 
as a tracer of star formation in obscured dusty regions, to place 
a firm upper bound on the energy from star formation, that is 
available to power the observed outbursts. 
Using NGC~4552's UV luminosity, 
calculated from photometric data at $2500$\AA (NED)
in Equation 1 from Kennicut (1998), we find an  unobscured  star 
formation rate of $0.09\,\Ms\,{\rm yr}^{-1}$ 
($0.14\,\Ms\,{\rm yr}^{-1}$) for a 
continuous (starburst) star formation model. This is an order of 
magnitude below that needed to power the observed outburst. Note that 
the unobscured star formation rate in the inner $20$\,pc nuclear region, 
calculated  using the UV flux measured by Maoz etal (2005) is very small, 
$< 0.0002-0.0003\,\Ms\,{\rm yr}^{-1}$. This is consistent with the 
high brightness temperature and flat spectral shape of NGC4552's radio 
emission that also argue against a dominant  compact nuclear starburst or 
collection of supernovae remnants in the galaxy's core (Nagar et al 2002). 

The TIR luminosity provides a tracer of star formation in dusty
regions, where the UV emission from young stars has been reprocessed 
to longer wavelengths by the surrounding dust. Since dust in 
elliptical galaxies tends to be centrally concentrated, and dust has
been identified in the inner $60$\,pc of NGC~4552 with HST observations
(van Dokkum \& Franx, 1995), this is likely a reasonable 
estimate for obscured star formation in the inner  
hundred parsecs of interest. We use the $60$ and 
$100\,\micron$ FIR flux densities for NGC~4552 from NED to calculate 
the FIR luminosity (Wise et al. 1993). We convert the FIR luminosity into the 
TIR luminosity using a correction factor ( $\lesssim 2$ for 
$F(60\,\micron)/F(100\,\micron)=0.3$; Helou \etal 1988). Then, using 
Equation 4 in  Kennicutt (1998), we find the star formation rate 
for obscured regions in NGC~4552 to be $\lesssim 0.03\,\Ms\,{\rm yr}^{-1}$.
Summing the UV and TIR star formation
rates, we find a firm upper limit for circumnuclear star formation of
$\lesssim 0.12-0.17\,\Ms\,{\rm yr}^{-1}$, more than a factor of five 
below that needed to power the outburst. These low upper limits on
current star formation in NGC~4552 are consistent with the $9.6$\,Gyr 
average age of the galaxy's stellar population determined from optical line 
ratios by Terlevich \& Forbes (2002).
Thus the outburst is most likely powered by NGC~4552's central AGN.  

VLBA measurements (Nagar \etal 2002) 
show that NGC~4552 has a core radio source with two-sided east-west
extended emission on $\sim 5$ milli-arcsecond scales.  
The radio spectrum is flat, with spectral index $\sim 0$ (Filho \etal
2004), and radio flux densities of $102.1$ and $99.5$\,mJy at $8.4$
and $5$\,GHz, respectively. Integrating this radio flux over 
the $10$ to $5000$\,MHz frequency band, we 
find a total central radio power of 
$L_{\rm radio} = 1.55 \times 10^{38}$\ergs.
Filho \etal (2004) argue that 
the properties of NGC~4552's radio
spectrum are most consistent with emission from 
accretion onto a $\sim 4 \times 10^{8}\Ms$ black hole plus a
self-absorbed compact jet or outflow. Our observation of shocks 
produced by outflows from a previous episode of nuclear activity  
supports this model.

Furthermore,  
the mechanical luminosity, 
$L_{\rm shock} \sim 3.3 \times 10^{41}$\ergs
(with factor $\sim 2$ uncertainty), found from our model fit to 
the observed shock in NGC~4552,  
is consistent with mechanical luminosities inferred for cavities 
produced by AGN outbursts in other systems.  
Using $18$ systems, ranging in size from
galaxy clusters to the elliptical galaxy M84 and containing
well defined X-ray surface brightness depressions (cavities),
Birzan \etal (2004) found a trend between the mechanical power 
$L_{\rm mech}$ needed to create the
X-ray cavities and the current central radio power $L_{\rm radio}$  
of the source, that they parameterized as a power law 
\begin{equation}
L_{\rm mech} = 
  10^{25 \pm 3}(L_{\rm radio})^{0.44 \pm 0.06}\, ,
\label{eq:birzan}
\end{equation}
albeit with large scatter. 
The error on the mechanical luminosity in Equation \ref{eq:birzan} 
reflects the several orders of magnitude uncertainty due to scatter in
their data.
From  Equation \ref{eq:birzan} , we expect the central radio source in 
NGC~4552 to produce 
outburst cavities  with mechanical power  
${\rm Log}\,L_{\rm mech} = 41.8 \pm 3$\ergs, where again the errors 
reflect the uncertainty due to scatter in the data around the trend
line. 
This is in good agreement with the 
mechanical power 
${\rm Log}\,L_{\rm shock} = 41.5 \pm 0.3$\ergs
found from our model fit to the observed shocks in NGC~4552, where the
errors in our observed value reflect the factor two uncertainty in the shock
model results.  

\section{The Nuclear Region}
\label{sec:nucleus}

Finally, we look briefly at the X-ray properties of the 
nuclear region by modeling the 
spectrum from a $1''.3$ circular region (NS) centered at 
($\alpha = 12^h35^m39.8^s$, $\delta = +12^\circ 33'22.9''$)  
containing the nuclear source (see Table \ref{tab:centralregs}). 
After local background subtraction 
using a concentric annular region with inner and outer radii of 
$1''.6$ and $3''.2$, respectively,  
the spectrum contains $1046$ net source counts in the $0.3-5$\,keV 
energy band.
An acceptable fit to the data ($\chi^2/{\rm dof} = 52/45$) was found 
using an absorbed power law model with moderate hydrogen absorption 
($n_H = 1.1 \pm 0.4 \times 10^{21}$\cms ) and  
steep ($\Gamma = 2.2 \pm 0.2$) photon index. The fits were improved 
and the photon index softened to $\Gamma = 1.7 \pm 0.2$ 
(see Table \ref{tab:nucfits}) by adding a 
second, thermal APEC component to the
model and fixing the hydrogen absorption for both thermal and powerlaw
components at the Galactic value ($n_H = 2.59 \times 10^{20}$\cms). 
For these two 
component models, we first allow the 
temperature, metal abundance and photon index to vary,
but find the abundance is poorly constrained. We then fix the 
metal abundance at 
$A=0.7\,\Zs$  from the best fit spectral model for the ring region 
$R$ of NGC~4552 (see the first line in Table \ref{tab:ringfits}),
 and find the fit is unchanged, although the errors on 
the photon index are modestly reduced. 
We checked the stability of this two component fit by 
allowing the absorbing column to vary and found $n_H$ consistent with 
the Galactic value.

In the two component model, we find a photon index 
$\Gamma \sim 1.7 \pm 0.1$ and temperature 
$kT = 1.04^{+0.21}_{-0.17}$\,keV. The power law 
photon index is consistent with those expected for low
luminosity AGNs (Terashima \& Wilson 2003; Filho \etal 2004). 
The temperature for the thermal
component is higher than that found for gas in any other region of the
galaxy, suggesting that we may be directly observing the reheating of
the galaxy ISM by the nuclear outburst.
We find intrinsic total $0.5-2$ and 
$2-10$\,keV X-ray fluxes in this model to be 
$5.1 \times 10^{-14}$\ergscm and 
$6.7 \times 10^{-14}$\ergscm, corresponding to $0.5 - 2.0$\,keV and 
$2-10$\,keV luminosities of $1.6 \times 10^{39}$\ergs and 
$2.1 \times 10^{39}$\ergs. 
The power law component dominates the emission in both energy bands, 
contributing $75\%$ and $97\%$ of the emission in the 
$0.5-2$  and $2 - 10$\,keV bandpasses, respectively. 
These results are in good agreement with the 
spectral results of Filho \etal (2004) for a circular 
region $2''.5$ in diameter surrounding the nucleus with data 
extracted from the same 
{\it Chandra} observation (Obsid 2072) used in this work.

Since the nuclear source is known to be variable at the 
$\sim 20\%$  level for the $2$ and $3.6$\,cm radio flux on 
timescales $\sim 1$\,yr (Nagar \etal 2002) and by a similar amount
in the UV at $2500$ and $3300$\AA\,    
on timescales of $\sim 2$ months (Maoz \etal 2005), 
it is interesting to look for X-ray variability.  
We examined the light curve for a $2''$ radius circular region centered 
on the nucleus 
($\alpha = 12^h35^m39.8^s$, $\delta = +12^\circ 33'22.9''$)
over the $0.3-10$\,keV energy range using $3$\,ks binning, 
and found that, on these timescales, the nuclear count rate 
varied by $\lesssim 2\,\sigma$ ($20 \%$) during the  
$52$\,ks observation.  
We then searched for variability on any scale using an 
application of the Gregory-Loredo algorithm (1992) on the 
unbinned events time series. We only found possible variability 
in the lowest order $m=2$ moment (${\rm Log(Odds Ratio)} = 36.9$, 
with the normalized odds ratio of $0.95$), consistent with a slow ($\lesssim
2\sigma$) increase in the mean signal over the duration of this observation.
 The Gregory-Loredo analysis failed to find 
any statistically significant evidence for variability on shorter 
timescales, i.e. in the $m > 2$ moments. Thus we find no statistical 
evidence in these data to support the claim by Xu \etal (2005), that 
was based on their ``visual inspection of the light curve'', for 
X-ray variability on $1$\,hr timescales.

\section{CONCLUSIONS}
\label{sec:conclude}

In this work we analysed a $54.4$\,ks {\it Chandra} observation 
of the elliptical galaxy NGC~4552 in the Virgo Cluster and found 
X-ray evidence for shocks in the inner region of the galaxy from 
a recent nuclear outburst. Two ring-like emission features, 
consistent with bipolar 
nuclear outflow cavities, are found in the X-ray images 
at $r \sim 1.3$\,kpc from NGC~4552's center.
The emission measure weighted gas temperature 
through the rings ($kT_{\rm proj} \sim 0.61 \pm 0.02$\,keV) and the 
shape of the surface brightness profile across the rim of the 
northern ring are consistent with a simple spherical
model for a Mach $1.7$ shock  
from a $1.4 \times 10^{55}$\,ergs nuclear outburst $\sim 1 - 2$\,Myr 
ago. The mechanical power carried in the shock, 
$L_{\rm shock} \sim 3.3 \times 10^{41}$\ergs, is consistent 
with that expected from the $L_{\rm mech}$-$L_{\rm radio}$ correlation
(Birzan \etal 2004) between the mechanical luminosity and total 
radio power for X-ray sources showing nuclear outflow activity.
One outburst of this magnitude every $\approx 18$\,Myr could  
balance the radiative cooling of NGC~4552's 
hot ISM. 

The X-ray spectrum for the $1''.3$ circular region containing the
nuclear point source is well fit by a two component power law + APEC 
model, with power law photon index $\Gamma = 1.7 \pm 0.1$ and gas 
temperature $kT=1.04^{+0.21}_{-0.17}$, in agreement with previous 
work by Filho \etal (2004).  The higher temperature found for gas in 
the nuclear region compared to that elsewhere in the galaxy suggests 
that we may be may be directly observing the reheating of the galaxy 
ISM in this region by the nuclear outbursts. 
 

\acknowledgements

This work is supported in part by NASA grant GO3-4176A  
and the Smithsonian Institution. 
This work has made use of the NASA/IPAC Extragalactic Database (NED)
which is operated by the Jet Propulsion Laboratory, California
Institute of Technology,  under contract with the National
Aeronautics and Space Administration. We wish to   
thank Maxim Markevitch for the use of his edge analysis codes and 
Arnold Rots for help in analysing the nuclear point source 
variability. 

\begin{small}

\end{small}

\begin{deluxetable}{ccccc}
\tablewidth{0pc}
\tablecaption{Spectral Analysis Regions for the Central 
$1.3$\,kpc of NGC~4552\label{tab:centralregs}}
\tablehead{
\colhead{Region }&\colhead{Type} &\colhead{Shape} & \colhead{Center}& 
\colhead{Dimensions} \\
 &  &   & RA, DEC   &  arcsec }
\startdata
R & source     & annular  & $12\,\,35\,\,39.9$, $12\,\,33\,\,23.4$ & $11$,$17$  \\
R & background & annular  & $12\,\,35\,\,39.9$, $12\,\,33\,\,23.4$ & $20$,$31$ \\
NS & source    & circular & $12\,\,35\,\,39.8$, $12\,\,33\,\,22.9$ & $1.3$ \\
NS & background & annular & $12\,\,35\,\,39.8$, $12\,\,33\,\,22.9$ & $1.6$,$3.2$ \\   
\enddata
\tablecomments{ WCS coordinates for the centers of the 
regions are J2000. Dimensions specified are radii for 
circular regions,  and (inner, outer) radii for annular regions.
Region R is also shown Figure \protect\ref{fig:ringimage}. 
 }
\end{deluxetable}

\begin{deluxetable}{ccccc}
\tablewidth{0pc}
\tablecaption{Spectral Models of NGC~4552's Rings\label{tab:ringfits}}
\tablehead{
\colhead{$kT$} & \colhead{$A$}&   
  \colhead{$K$}&  \colhead{$\Lambda$}&\colhead{$\chi^2/{\rm dof}$} \\
keV  & $\Zs$ &$10^{-5}$\,cgs &$10^{-23}$erg\,cm$^{3}$s$^{-1}$ & }
\startdata
$^\dagger\,0.64 \pm 0.02$ &$0.7^{+0.4}_{-0.2}$&$5.8$ &$1.7$ &$45.8/55$  \\
 $0.65 ^{+0.02}_{-0.03}$  &$0.4$ &$9.1$  & $1.1$   & $55/56$ \\
 $0.64^{+0.03}_{-0.02}$   &$0.5$  &$7.6$  & $1.3$  & $48.8/56$ \\
 $0.64^{+0.03}_{-0.02}$   &$0.6$  &$6.6$  & $1.5$  & $46.4/56$ \\
 $0.64 \pm 0.02$          &$0.8$  &$5.1$  & $1.9$  & $46/56$ \\
 $0.64 \pm 0.02$          &$1.0$  &$4.2$  & $2.4$  & $47.4/56$ \\
\enddata
\tablecomments{Spectral model parameters for the rings 
from region R (see Table \protect\ref{tab:centralregs} and 
Figure \protect\ref{fig:ringimage}), using an absorbed 
single temperature APEC model fit over the $0.3-3$\,keV energy range. 
The spectrum contains $1840$ net source counts. The hydrogen absorption 
column is fixed at the Galactic value 
($n_{\rm H}=2.59 \times 10^{20}$\cms).
$^\dagger$\,The first row denotes the best fit model with temperature and
abundance free to vary, while following rows hold the abundance fixed.
Errors are $90\%$ confidence limits.
}
\end{deluxetable}

\begin{deluxetable}{ccccc}
\tablewidth{0pc}
\tablecaption{Spectral Models of the Nuclear Region of NGC~4552\label{tab:nucfits}}
\tablehead{
\colhead{$n_{\rm H}$} &\colhead{$kT$} &\colhead{$A$}  
  & \colhead{$\alpha$}&\colhead{$\chi^2/{\rm dof}$} \\
 $10^{20}$\cms& keV & $\Zs$ &  & }
\startdata
$11 \pm 4$ &\nodata  &\nodata  &$2.2 \pm 0.2$ &$51.7/45$  \\
$2.59^f$  &$1.04^{+0.13}_{-0.17}$ & $0.5^{+0.9}_{-0.4}$ &
$1.7^{+0.2}_{-0.3}$ & $35.6/43$  \\
$2.59^f$  &$1.04^{+0.21}_{-0.11}$ & $0.7^f$ &
$1.7 \pm 0.1$ & $35.7/44$  \\
$< 6.7$   & $1.04^{+0.13}_{-0.18}$ & $0.7^f$ & $1.7 \pm 0.2$  & $35.6/43$  \\
\enddata
\tablecomments{ The spectrum contains $1046$ net source counts in 
region NS, specfied in Table \protect\ref{tab:centralregs}, over
the $0.3-5$\,keV band used for these fits. Superscript $f$ denotes a 
fixed parameter. Errors are $90\%$ confidence limits. 
 }
\end{deluxetable}
\vfill
\eject
\end{document}